# New Superconductor $(Na_{0.25}K_{0.45})$ $Ba_3Bi_4O_{12}$: A First-principles Study

Ali M S[1] Aftabuzzaman M[1] Roknuzzaman M[2] Rayhan M A[3] Parvin F[3] Ali M M[4] Rubel M H K[4] and Islam A K M A[3, 5*]

[1]Department of Physics, Pabna University of Science and Technology, Pabna 6600, Bangladesh
[2]Department of Physics, Jessore University of Science and Technology, Jessore 7408, Bangladesh
[3]Department of Physics, Rajshahi University, Rajshahi 6205, Bangladesh
[4]Center for Crystal Science and Technology, University of Yamanashi, 7-32 Miyamae Kofu 400-8511 Japan
[5]International Islamic University Chittagong, 154/A College Road, Chittagong-4203, Bangladesh

**Abstract**
A new superconductor $(Na_{0.25}K_{0.45})Ba_3Bi_4O_{12}$, having an *A*-site-ordered double perovskite structure, with a maximum $T_c \sim 27$ K has very recently been discovered through hydrothermal synthesis at 593 K. The structural, elastic, electronic, and thermal properties of the new synthesized compound have been investigated theoretically. Here we have employed the pseudo-potential plane-wave (PP-PW) approach based on the density functional (DFT) theory, within the generalized gradient approximation (GGA). The elastic constants ($C_{ij}$), Pugh`s ratio, Cauchy`s pressure and other elastic parameters are derived and analyzed using energy strain method for the first time. We have discussed the bonding nature in the light of the electronic valence charge density. Both electron and hole-like Fermi surfaces are present in the compound under study which indicate the multiple-band nature of $(Na_{0.25}K_{0.45})Ba_3Bi_4O_{12}$. The compound is indicated to be a strongly coupled superconductor which is based on the estimated e-ph coupling constant. The thermodynamic properties such as bulk modulus, Debye temperature, heat capacities and volume thermal expansion coefficient at elevated temperature and pressure are calculated and analyzed for the first time by using quasi-harmonic model.
.

## 1. Introduction

A new superconductive bismuth oxide with a double perovskite-type structure of $(Na_{0.25}K_{0.45})Ba_3Bi_4O_{12}$ has been synthesized by Rubel *et al* [1]. Previously, the highest $T_c$ belonged to the perovskite superconductor family based on the bismuthates, $BaBiO_3$ [2]. As early as 1975 Sleight *et al* [3] first discovered the superconductivity in $BaPb_{1-x}Bi_xO_3$ with a maximum $T_c \sim 13$ K. A new superconductor $Ba_{1-x}K_xBiO_3$ (BKBO) with a higher $T_c \sim 30$ K was reported by Cava *et al* [4, 5] and Matheiss *et al* [6] by doping K at Ba site in $BaBiO_3$. In fact the electronic properties of (BKBO) have been studied extensively to investigate metalicity at ambient conditions by several authors [7-9].

The semiconductor $BaBiO_3$ crystallizes in a monoclinic distorted order as a perovskite structure [2], where the $BiO_6$ octahedra are tilted at a lower angle, in contrast to the 180° orientation in the perfect structure. A series of compounds can be found by substituting K, Pb, and Sr atoms in the $BaBiO_3$ compound to form $Ba_{1-x}K_xBiO_3$, $Ba_{1-x}Pb_xO_3$, and $Sr_xK_{1-x}BiO_3$ which induces a symmetry change with the appearance of superconductivity.

The noncuprate superconductor, $Ba_{1-x}K_xBiO_3$ (BKBO), has a simple perovskite-type structure in which the *A*-sites are occupied by $K^+$ and $Ba^{2+}$ ions and the *B*-sites are occupied by bismuth ions in mixed valence states with a $BiO_6$ octahedron. A superconductive perovskite-type oxide $(Ba_{0.75}K_{0.14}H_{0.11})BiO_3 \cdot nH_2O$, with $T_c = 8$ K, fabricated by hydrothermal synthesis at 180 °C has previously been reported [10]. The crystal structure of this reported compound was determined to be double perovskite. Zhang *et al* [11] also reported the hydrothermal synthesis of a double perovskite with the structure $Ba_{1-x}K_xBi_{1-}$ $_y$Na$_y$O$_3$. However, no crystal structural refinement was performed; it also did not show $T_c$ even as low as 2 K [1].

Thus efforts were made to understand the effect of the synthesis conditions and crystal structures on the emergence of superconductivity in Bi-based double perovskite oxides [1]. These efforts finally led to the discovery of the double perovskite superconductor. At the time of discovery structural properties have been determined and some attempts to examine theoretically metallic behavior of (Na$_{0.25}$K$_{0.45}$)Ba$_3$Bi$_4$O$_{12}$ [1]. But experimental and theoretical information are not available yet as on elastic, electronic (Fermi surface, electronic charge density), and thermodynamic properties of the new superconductor. In this paper we have calculated these properties for the first time. The mechanical properties such as independent elastic constants, mechanical stability, Young's modulus, bulk modulus, shear modulus, Cauchy's pressure, Poisson's ratio, elastic anisotropy, and Peierls stress are calculated. The band structure, density of states, Fermi surface and total charge density distribution in the (101) plane have been calculated. The thermodynamic properties such as bulk modulus, Debye temperature, heat capacities and volume thermal expansion coefficient at elevated temperature and pressure are calculated and analyzed for the first time by using quasi-harmonic model.

## 2. Theoretical Methods

The calculations have been carried out using CASTEP code which is based on the density functional theory (DFT) [12-15]. Furthermore, the ultrasoft pseudopotential formalism of Vanderbilt [16] is used to simulate the interactions of valence electrons with ion cores, and the electron wave function is expanded in plane waves up to an energy cutoff of 410 eV for all calculations. The exchange–correlation energy is evaluated using the GGA of the Perdew–Burke–Ernzerhof for solids (PBEsol) formalism [15], which is dependent on both the electron density and its gradient at each space point. To search the ground state, a quasi-Newton (variable-metric) minimization method using the Broyden–Fletcher–Goldfarb–Shanno (BFGS) update scheme [17] is utilized, which provides a very efficient and robust way to explore the optimizing crystal structure with a minimum energy. The plane wave basis set cut-off is set as 410 eV, and for the sampling of the Brillouin zone [18], an 8×8×8 Monkhorst–Pack mesh is employed [19]. Geometry optimization is achieved using convergence thresholds of $10^{-5}$ eV/atom for the total energy, 0.03 eV/Å for the maximum force, 0.05 GPa for the maximum stress and $10^{-3}$ Å for maximum displacement. By utilizing a set of homogeneous deformations with a finite value under linear proportion, the resulting stress can be calculated with respect to the optimized crystal structure [20]. The elastic coefficients can be finally determined through a linear fit of the calculated stress as a function of strain, where four strain amplitudes are used with the maximum value of 0.3%.

The thermodynamic properties have been studied within the quasi-harmonic Debye model implemented in the Gibbs program [21]. The detailed description of the model can be found in literature [22-25]. Through this model, one could calculate the thermodynamic parameters including the bulk modulus, thermal expansion coefficient, specific heats, and Debye temperature etc. at any temperatures and pressures using the DFT calculated $E$-$V$ data at $T = 0$ K, $P = 0$ GPa and the Birch-Murnaghan EOS [26].

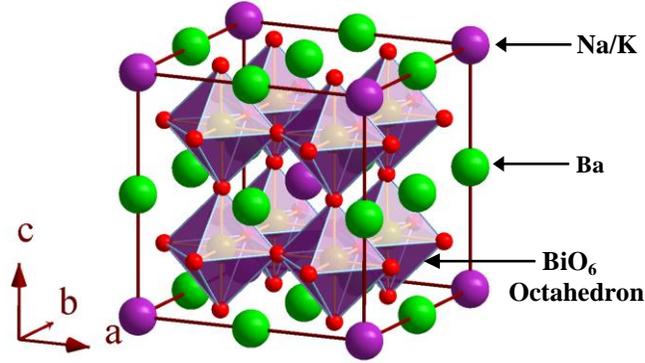

**Figure 1.** Crystal structure of A′Ba$_3$B$_4$O$_{12}$-type perovskite (A′ = Na/K).

## 3. Results and Discussion

### 3.1. Structural properties

The lattice constants and atomic positions of (Na$_{0.25}$K$_{0.45}$)Ba$_3$Bi$_4$O$_{12}$ have been optimized as a function of normal stress by minimizing the total energy. The optimized structure is shown in Fig. 1. Experimental studies have established that (Na$_{0.25}$K$_{0.45}$) Ba$_3$Bi$_4$O$_{12}$ crystallizes in cubic structure with space group *Im-3m* (#229) and has 40 atoms in one unit cell. The calculated ground state lattice constant *a* within GGA is listed in Table 1. The calculated value agrees reasonably with the experimental result [1]. In the crystal structure Na and K randomly occupy leaving some vacancies while Bi fully occupies and forms corner–sharing BiO$_6$ octahedral network.

**Table 1.** Calculated lattice constant *a* (Å), elastic constants $C_{ij}$ (GPa), bulk moduli *B* (GPa), shear moduli *G* (GPa), Young's moduli *E* (GPa), Poisson's ratio *v*, elastic anisotropic factor *A*, Burger's vector *b* (Å), interlayer distance *d* (Å), and Peierls stress $\sigma_p$ for (Na$_{0.25}$K$_{0.45}$)Ba$_3$Bi$_4$O$_{12}$ at *P* = 0 GPa and *T* = 0 K.

| *a* | $C_{11}$ | $C_{12}$ | $C_{44}$ | *B* | *G* | *E* | *v* | *A* | *b* | *d* | $\sigma_p$ |
|---|---|---|---|---|---|---|---|---|---|---|---|
| 8.6277 | 223 | 48 | 53 | 107 | 65 | 162 | 0.25 | 0.61 | 8.6277 | 4.314 | 2.63 |
| 8.5493[a] | | | | | | | | | | | |

[a]Ref. [1]

### 3.2. Elastic properties

The elastic moduli require the knowledge of the derivative of energy as a function of lattice strain. To study the elastic properties of (Na$_{0.25}$K$_{0.45}$)Ba$_3$Bi$_4$O$_{12}$ at *T* = 0 K and *P* = 0 GPa, the elastic constants $C_{ij}$, bulk modulus *B*, shear modulus *G*, Young's modulus $E = 9BG/(3B + G)$, and Poisson's ratio $v = (3B - E)/6B$, and anisotropic factor *A* have been calculated for the first time and are presented in Table 1. Theoretical details on the calculation of these constants can be found elsewhere [27, 28].

It has been suggested that the Born conditions are valid only for the stability analysis of an unstressed lattice and not for the stressed lattice [29, 30]. The stability criteria have been formulated in terms of the elastic stiffness coefficients ($C_{ij}$) which govern the proper stress–strain relations at finite strain by considering both the internal energy and the external work done during deformation [29, 30]. The Born stability criteria, which define the mechanical stability of a lattice, are typically formulated in terms of the $C_{ijkl}$. In this case, they are valid only in the limit of vanishing of pre-stress. Several studies [31, 32] have demonstrated that the appropriate stability criteria for a stress lattice are those which are formulated in terms of the stress-strain coefficients ($C_{ijkl}$) and hence are based on enthalpy considerations. Under hydrostatic pressure, the three stability criteria for a cubic crystal are

$C_{11}+2C_{12}> 0$, $C_{44}>0$, $C_{11}-C_{12}>0$

These are referred to as spinodal, shear, and Born criteria, respectively. The spinodal criterion is equivalent to requiring that the bulk modulus be positive.

The bulk modulus is defined as the resistance to volume change of a body; the shear modulus reflects the resistance to shape change of a body. The conditions for mechanical stability in a cubic crystal leads to the restrictions on the elastic constants that $B$, $C_{11}-C_{12}$ and $C_{44}$ should be positive [33]. The elastic constants presented in Table 1 obey these stability conditions showing that $(Na_{0.25}K_{0.45})Ba_3Bi_4O_{12}$ is mechanically stable.

Cracks in crystals are directly related to the anisotropy of thermal and elastic properties. The Zener elastic anisotropy of crystal, defined by the ratio $A = 2C_{44}/(C_{11} - C_{12})$ [34], yields a value of 0.61 for $A$. The factor $A = 1$ represents complete isotropy, while value smaller or greater than this measures the degree of anisotropy. Till date, no experimental or theoretical data for the elastic stiffness are available in the literature for comparison with our theoretical results. Cubic symmetry requires three independent elastic constants to describe its elastic response. Unidirectional elastic constant $C_{11}$ which relates to $a$-direction is about 4.21 times higher than the pure shear elastic constant $C_{44}$.

The ductile-brittle nature of materials can be discussed in terms of elastic constants of the relevant material. If the Cauchy's pressure ($C_{12}-C_{44}$) is negative (positive), the material is expected to be brittle (ductile) [35]. In the present case this value is negative which indicates that $(Na_{0.25}K_{0.45})Ba_3Bi_4O_{12}$ is brittle. Another index of ductility is Pugh's ratio [36] and a material behaves in a ductile manner, if $G/B < 0.5$, otherwise it should be brittle. Thus the value of 0.61 for $(Na_{0.25}K_{0.45}) Ba_3Bi_4O_{12}$ indicates that it is just above the border of brittleness.

Peierls stress is the force needed to move a dislocation within a plane of atoms in the unit cell. The movement of a dislocation in a glide plane of the double perovskite can be predicted using Peierls stress ($\sigma_P$) [37]. The initial stress value required to initiate such movement is expressed through shear modulus $G$ and Poisson ratio $\nu$ as follows:

$$\sigma_P = \frac{2G}{1-\nu}\exp\left(-\frac{2\pi d}{b(1-\nu)}\right) \tag{1}$$

where $b$ is the Burgers vector and $d$ is the interlayer distance between the glide planes.. The calculated Burger's vector $b$, interlayer distance $d$, and the resulting Peierls stress $\sigma_p$ for $(Na_{0.25}K_{0.45}) Ba_3Bi_4O_{12}$ are presented in Table 1.

The estimated Peierls stress data for the double perovskite is 2.63 which may be compared with those of inverse perovskites $Sc_3InX$ ($X$ = B, C, N) [38] and a selection of MAX phases $Ti_2AlC$, $Cr_2AlC$, $Ta_2AlC$, $V_2AlC$, and $Nb_2AlC$ for which $\sigma_p$ are in the ranges 0.7 to 0.98 (GPa) [38, 39]. The reported $\sigma_p$ for rocksalt binary carbide TiC is 19.49 [39], showing the sequence, $\sigma_p$ (selected inverse perovskites and MAX) < $\sigma_p$ (double perovskite) ≪ $\sigma_p$ (binary carbide). It is thus clear that in the selected MAX phases dislocations can move, but this is not the case for the binary carbide. Since the superconducting double perovskite studied here exhibit an intermediate value of $\sigma_p$, approximately 3 times larger than in MAX phases, dislocation movement may still occur here, but not as easily as in MAX phases.

*3.3. Electronic properties*

The calculated electronic band structures along the high symmetry directions in the Brillouin zones together with the total density of states of $(Na_{0.25}K_{0.45})Ba_3Bi_4O_{12}$ are shown in Fig. 2. It is observed from the band structure, which is very much similar to that in ref. [1], that the compound under study is metallic because a number of bands are overlapping at the Fermi level. At the Fermi level, a strong hybridization is found between Bi-6s and O-2p orbitals. But O-2p contributes largely than Bi-6s orbital which agrees well with Rubel *et. al.* [1]. It can be seen from the figure of DOS that the Fermi level is located at the vicinity of the DOS peak, which leads large DOS at the Fermi level with values of 5.2 states/eV for $(Na_{0.25}K_{0.45})Ba_3Bi_4O_{12}$.

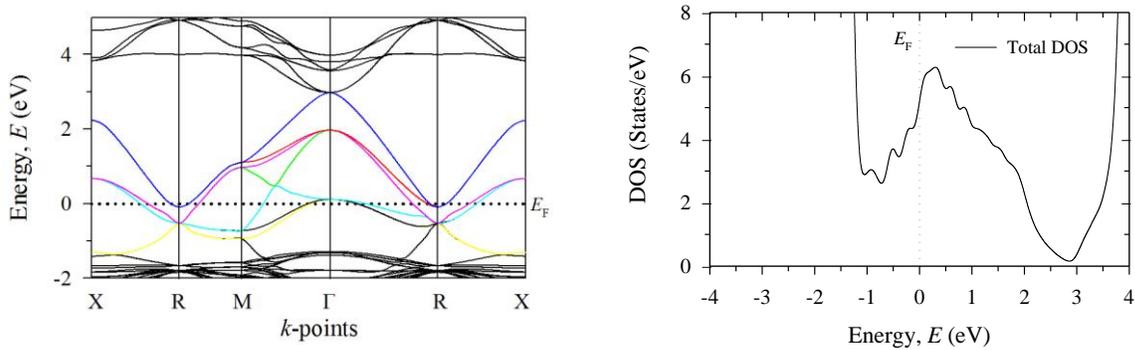

**Figure 2.** Electronic band structure and total DOS of $(Na_{0.25}K_{0.45})Ba_3Bi_4O_{12}$.

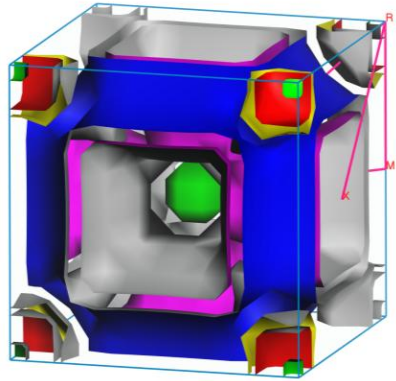

**Figure 3.** Fermi surface of $(Na_{0.25}K_{0.45})Ba_3Bi_4O_{12}$.

Fig. 3 shows the Fermi surface of the double perovskite $(Na_{0.25}K_{0.45})Ba_3Bi_4O_{12}$ for the bands crossing the Fermi level (colored lines). The topology shows that at the $\Gamma$-point there is a hole-like Fermi surface. There is also a hole pocket around X-point which is connected with another hole-like Fermi surface surrounding $\Gamma$-point. At the corner of the Brillouin zone three electron-like Fermi surfaces are present around R-point. So, it is seen that both electron and hole-like Fermi surfaces are present in the compound under study which indicate the multiple-band nature of $(Na_{0.25}K_{0.45})Ba_3Bi_4O_{12}$.

In order to understand the distribution of the total electronic charge density maps of $(Na_{0.25}K_{0.45})Ba_3Bi_4O_{12}$ compound, the valence electronic charge density maps (in the units of e/Å$^3$) have been depicted in Fig. 4 along (1 0 1) crystallographic plane. The scale shows behind the right side of Fig. 4 represent the intensity of electron density. The blue color shows the light density of electron whereas the red color shows high density of electron. The O-Bi bonds are very strong, which coincides with the strong hybridization between O-2$p$ and Bi-6$s$ electrons in DOS (not shown). Also we have found from structural properties that O-Bi formed octahedral structure with strong ionic bond. We should emphasize that the charge density distribution is essentially spherical around all the atoms which shows ionic nature. The O atom is more electronegative than Bi, Ba and K/Na atoms, as one can clearly see that the charge accumulates more near O along the bonds and the charge around O is uniformly distributed. The electronic charge densities in all crystallographic planes are the same showing its isotropic nature.

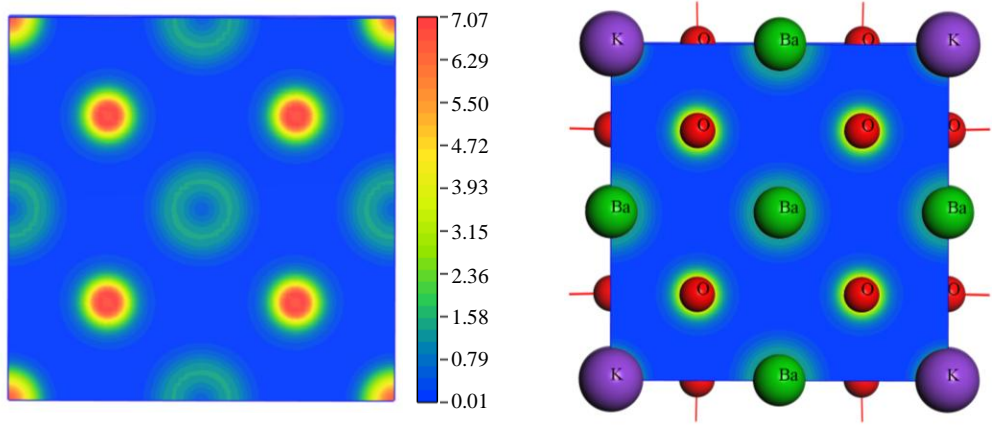

**Figure 4.** Electronic charge density of $(Na_{0.25}K_{0.45})Ba_3Bi_4O_{12}$.

*3.4. Electron-phonon coupling constant*

The electron-phonon coupling constant $\lambda$, can be estimated using McMillan's equation [40]

$$\lambda = \frac{1.04 + \mu^* \ln\left(\frac{\Theta_D}{1.45 T_c}\right)}{(1-0.62\mu^*)\ln\left(\frac{\Theta_D}{1.45 T_c}\right) - 1.04} \qquad (2)$$

where $\mu^*$ represents the repulsive screened Coulomb potential. Using our calculated Debye temperature $\Theta_D = 379$ K, measured $T_c = 27$ K [1] and $\mu^* = 0.13$ in above equation we get $\lambda = 1.27$ for $(Na_{0.25}K_{0.45})Ba_3Bi_4O_{12}$ which implies that the compound is typically a strongly coupled superconductor.

*3.5. Thermal properties*

It has been shown by several authors that various thermodynamic properties can be successfully obtained through Debye quasi-harmonic model which is gauged from comparison with the available experimental data (for example see ref. [41]). We apply the quasi-harmonic approximation to investigate the thermodynamic properties of $(Na_{0.25}K_{0.45})Ba_3Bi_4O_{12}$, in which the nonequilibrium Gibbs function $G^*(V; P, T)$ can be written as [21]:

$$G^*(V; P, T) = E(V) + PV + A_{vib}(\Theta_D(V); T), \quad (3)$$

Where $E(V)$ is the total energy per unit cell, $PV$ corresponds to the constant hydrostatic pressure condition and $A_{vib}(\Theta_D(V); T)$ is the vibrational Helmholtz free energy which given by

$$A_{vib}(\Theta_D(V); T) = nkT\left[\frac{9\Theta_D}{8T} + 3\ln\left(1 - e^{-\Theta_D/T}\right) - D\left(\frac{\Theta_D}{T}\right)\right] \quad (4)$$

where $n$ is the number of atoms in per formula unit, $k$ is the Boltzmann constant, $D(\Theta_D/T)$ is the Debye integral and $\Theta_D$ represents the Debye temperature. The thermal equation of state $V(P,T)$ and the chemical potential $G(P,T)$ can be obtained by minimizing the nonequilibrium Gibbs function with respect to volume $V$. Other macroscopic properties as a function of pressure $P$ and temperature $T$ can also be derived from standard thermodynamic relations [21].

The bulk modulus, Debye temperature, specific heats, and volume thermal expansion coefficient at different temperatures and pressures are evaluated. For this we utilized $E$–$V$ data obtained from the third-order Birch–Murnaghan equation of state [26] using zero temperature and zero pressure equilibrium values of $E_0$, $V_0$, $B_0$, based on DFT method. The thermal properties are determined in the temperature range from 0 to 1000 K, where the quasi-harmonic Debye model are seen to be valid. The pressure effect is also studied in the 0 - 35 GPa range.

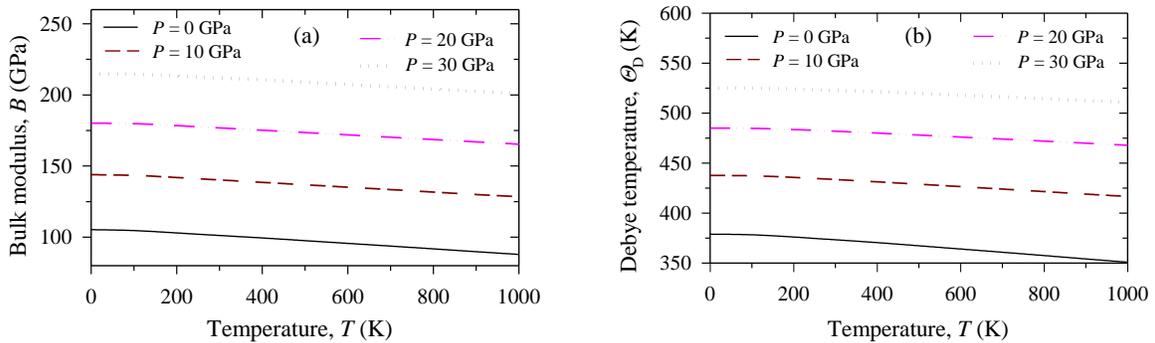

**Figure 5.** The temperature dependence of (a) bulk modulus and (b) Debye temperature of $(Na_{0.25}K_{0.45})Ba_3Bi_4O_{12}$.

The temperature dependence of bulk modulus of $(Na_{0.25}K_{0.45})Ba_3Bi_4O_{12}$ is shown in Fig. 5(a). It is seen that the bulk modulus is nearly constant from 0 to 100 K and then decreases linearly with increasing temperature. It is also found that when the temperature is constant, bulk modulus increases with applied pressure. Again, the concept of Debye temperature is very important in determining the thermodynamic properties of materials. It is basically a measure of vibrational response of the material and is therefore intimately connected with properties like specific heat and thermal expansion. Fig. 5(b) shows the variation of Debye temperature $\Theta_D$ for $(Na_{0.25}K_{0.45})Ba_3Bi_4O_{12}$ with temperature. It is clearly observed from Fig. 5(b) that $\Theta_D$ decreases linearly with the increase in temperature whereas it increases with the increase in pressure. The variation of $\Theta_D$ with pressure and temperature reveals that the thermal vibration frequency of atoms in the double perovskite compound changes with pressure and temperature. The Debye temperature and bulk modulus, at $P = 0$ and $T = 0$ are found to be 379 K and 107 GPa which are in good agreement with the values of 315 K and 106 GPa computed using our elastic constant data [42-44]. This might be an indication that the quasi-harmonic Debye model is a reasonable alternative to account for the thermal effects with no expensive task in terms of quantum-mechanical phononic model.

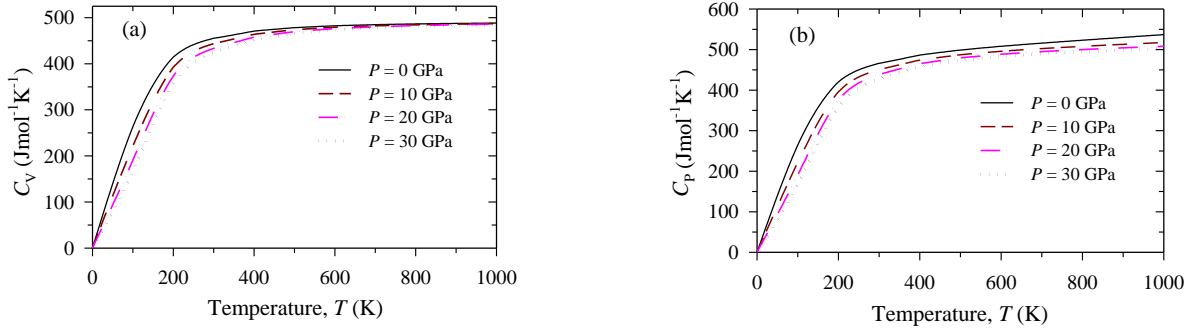

**Figure 6.** The temperature dependence of (a) specific heat at constant volume and (b) specific heat at constant pressure of $(Na_{0.25}K_{0.45})Ba_3Bi_4O_{12}$

The knowledge of the heat capacity of a substance provides essential insight into its vibrational properties. The temperature dependence of the heat capacities is presented in Fig. 6. It is observed that as the temperature increases, the heat capacity $C_V$ increases. At high temperature ($T > 300$ K), the calculated $C_V$ converges to a constant $C_V = 3nNk_B = 488.4$ J/mol.K), which is in agreement with the law of Dulong and Petit. At sufficiently low temperature, $C_V$ is proportional to $T^3$ [45]. The same phenomena is observed in case of specific heat at constant pressure $C_P$.

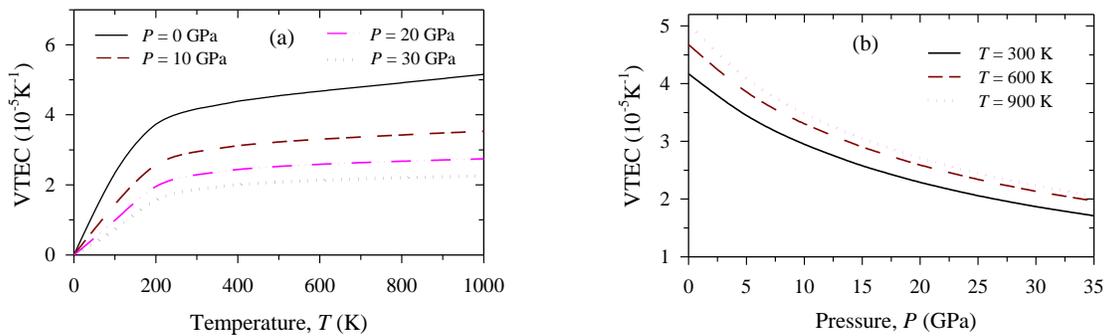

**Figure 7.** Volume thermal expansion coefficient, VTEC ($\alpha_v$) of $(Na_{0.25}K_{0.45})Ba_3Bi_4O_{12}$ as a function of (a) temperature, and (b) pressure.

Finally, the variations of the volume thermal expansion coefficient $\alpha_v$ with temperature and pressure for $(Na_{0.25}K_{0.45})Ba_3Bi_4O_{12}$ are presented in Fig. 7. The volume thermal expansion coefficient, $\alpha$ reflects the temperature dependence of volume at constant pressure: $\alpha = \frac{1}{V}\left(\frac{dV}{dT}\right)_P$. The temperature dependence of $\alpha_v$ at different pressures is displayed in Fig. 7(a). The coefficient $\alpha_v$ increases rapidly as the temperature increases upto ~250 K and then approaches to a slow increase at higher temperature, whereas for $P > 10$ GPa, it tends to be constant. Fig. 7 (b) shows the pressure dependence of $\alpha_v$ at three different temperatures. For a given temperature, the coefficient $\alpha_v$ sharply decreases with the increase of pressure. As there are no experimental results or other theoretical calculations for the thermal properties of $(Na_{0.25}K_{0.45})Ba_3Bi_4O_{12}$, our calculations may provide a helpful reference for further study.

## 4. Conclusion

The newly discovered superconductor $(Na_{0.25}K_{0.45})Ba_3Bi_4O_{12}$ has been subjected to theoretical investigation on structural, elastic, electronic, and thermal properties. The calculated elastic constants ($C_{ij}$) satisfy the Born mechanical stability criterion. The brittle nature of the newly synthesized compound was investigated using Pugh`s ratio and Cauchy`s pressure criteria. The estimated Peierls stress shows that dislocation movement should not happen here as easily as in the cases such as some selected MAX phases.

Strong hybridization between O-$2p$ and Bi-$6s$ orbitals was observed in the band structure, DOS and electron charge density map. The charge density distribution is essentially spherical around all the atoms showing ionic nature. The multiple-band nature of the compound is evident from the existence of both electron and hole-like Fermi surfaces. The estimated electron-phonon coupling constant ($\lambda = 1.27$) implies that the compound is typically a strongly coupled superconductor.

The thermodynamic properties such as bulk modulus, Debye temperature, heat capacities and volume thermal expansion coefficient of the double perovskite as a function of temperature and pressure have been evaluated and analyzed for the first time by using quasi-harmonic model.